\documentclass[fleqn,usenatbib]{mnras}

\usepackage{newtxtext,newtxmath}
\usepackage[T1]{fontenc}

\DeclareRobustCommand{\VAN}[3]{#2}
\let\VANthebibliography\thebibliography
\def\thebibliography{\DeclareRobustCommand{\VAN}[3]{##3}\VANthebibliography}

\usepackage{graphicx}	% Including figure files
\usepackage{amsmath}	% Advanced maths commands
\usepackage{booktabs}

\newcommand{\lp}{\left(}
\newcommand{\rp}{\right)}

\title[Constraints on the UHECR output of GRBs]{Constraints on the ultra-high energy cosmic ray output of gamma-ray bursts}

\author[E. Moore et al.]{
E. Moore,$^{1,2}$
B. Gendre,$^{1,2}$\thanks{E-mail: bruce.gendre@gmail.com}
N. B. Orange,$^{3,4}$
and F. H. Panther$^{1,2}$\\
$^{1}$Department of Physics, University of Western Australia, Crawley WA 6009, Australia\\
$^{2}$OzGrav: The ARC Centre of Excellence for Gravitational-wave Discovery\\
$^{3}$OrangeWave Innovative Science, LLC, Moncks Corner, SC 29461, USA\\
$^{4}$Etelman Observatory Research Center, University of the Virgin Islands, St. Thomas 00802, USVI, USA
}

\date{Accepted XXX. Received YYY; in original form ZZZ}

\pubyear{2024}

\begin{document}
\label{firstpage}
\pagerange{\pageref{firstpage}--\pageref{lastpage}}
\maketitle

\begin{abstract}
Ultra-high energy cosmic rays are the most extreme energetic particles detected on Earth, however, their acceleration sites are still mysterious. We explore the contribution of low-luminosity gamma-ray bursts to the ultra-high energy cosmic ray flux, since they form the bulk of the nearby population.  We analyse a representative sample of these bursts detected by BeppoSAX, INTEGRAL and Swift between 1998-2016, and found that in order to reconcile our theoretical flux with the observed flux, these bursts should accelerate at most $10^{-13}$ M$_\odot$ of ultra-high energy cosmic rays.
\end{abstract}


%%%%%%%%%%%%%%%%%%%%%%%%%%%%%%%%%%%%%%%%%%%%%%%%%%

%%%%%%%%%%%%%%%%% BODY OF PAPER %%%%%%%%%%%%%%%%%%

\section{Introduction}

Ultra-high energy cosmic rays (UHECRs) are the most energetic, yet elusive particles in the universe whose origins remain an outstanding problem in physics \citep{hill84,nag00, bat19}. By definition, these particles have energies $\geq 10^{18}$ eV. Numerous studies have been undertaken to improve the spectral resolution of the highest energy range ($\geq 10^{18}$ eV) of the cosmic ray all-particle spectrum \citep[see e.g.][]{abu01, abb04, abr08, aab16, alf17, aab20}. These studies first confirmed that UHECRs are located below a steepening in the energy spectrum of the cosmic rays, known as the ``ankle'' \citep{hill05}. They also confirmed the presence of a horizon known as the Greisen-Zatsepin-Kuzmin (GZK) horizon \citep{gre66, zat66, abb08}. That horizon is caused by the interaction of the cosmic microwave background (CMB) radiation and the cosmic ray protons, that results in a loss of energy of the latter, and impacts particles with energies $\geq 5 \times 10^{19.5}$ eV. Despite the GZK horizon, it is now well-established that the unknown sources of UHECRs are extra-galactic \citep{bat19}. It is also now being investigated if this steepening may be a result of limitations in the maximum obtainable energy from acceleration sites \citep[e.g.][]{aab17}.

Gamma-ray bursts (GRBs) are the most powerful and luminous extra-galactic transients in the universe \citep{kle73, zha18}. The fireball model \citep{cav78, ree92, mes97, pan98} is the leading model of GRB emission, wherein a central engine (likely a black hole) injects relativistic shells of plasma that collide and produce the gamma-rays observed in the prompt emission. Shell collisions within the burst provide ideal conditions for particle acceleration via the Fermi mechanism \citep{bel78, bla78}, hence GRBs have long been speculated as promising sources of UHECRs \citep{mil95, vie95, wax95}.

Recently, a high-energy photon thought to originate from the interaction of an UHECR emitted by GRB980425 has reinvigorated interest into the possibility that some high-energy protons are indeed due to GRBs \citep{mir22b, mir22}. GRB980425 is a representative of the low-luminosity GRB (LLGRB) population \citep{lia07, der17}, which have been theoretically explored as potential sources of UHECRs \citep[see e.g.][]{mur06, wan07, mur08,liu11, zha18b, bon19, rud22}; although see \citet{sam19, sam20}. Hence it is tempting to see if the observed rate of such events can account for the production of some or all UHECRs. 

In this paper, we use a sample of LLGRBs from \cite{der17} to determine their occurrence rate in the local universe. From that rate, we infer the production rate of UHECRs from LLGRBs and compare our result with observations.

This paper is organised as follows. In Section \ref{sec:data}, we describe the data set used in our statistical study, and in Section \ref{sec:results} we outline our method for computing the rates of LLGRBs and UHECRs. In Section \ref{sec:llgrb}, we discuss the nature of LLGRBs, and provide constraints on the number of UHECRs that can be accelerated in the jet by comparing our derived UHECR flux to the detected flux. In Section \ref{sec:discussion} we investigate potential suppression mechanisms that could bias the results, and discuss the implications our results have on the contents of a GRB jet. For our analysis, we assume a flat $\Lambda$-CDM cosmology with $H_0 = 67.66$ km s$^{-1}$ Mpc$^{-1}$ and $\Omega_M = 0.30966$ \citep{pla18}.

\section{Data}
\label{sec:data}

So far, only a tentative association of an UHECR has been found for GRB980425 \citep{mir22b, mir22}. This event is one of the closest GRBs ever recorded on Earth \citep{gal98}, but also one of the most under-luminous to date \citep{ama06}. \citet{der17} has shown that this burst was representative of a whole population of faint events that are not visible at large ($z \gtrsim 1$) redshifts, but are overabundant in the local Universe. We considered that this population of nearby events would be the principle source of UHECRs produced by GRBs. Thus, we use a sample of bursts similar to GRB980425 from \citet{der17} that were selected by their X-ray properties, which act as a proxy for the total fireball energy measurement. We rejected from that sample events with no known redshift, as we study the occurrence rate of these events corrected for the variation in detection distances. The final sample comprises of 41 LLGRBs with known redshift. They are listed in Table \ref{tab:2}, with some of their properties.

\begin{table}
\begin{tabular}{@{}lcll@{}}
\toprule
GRB & $E_{\text{iso}}$ {[}$10^{52}$ erg{]} & $z$ & Rate {[}Mpc$^{-3}$ yr$^{-1}${]} \\ \midrule
GRB980425 & $(1.3\pm0.2)\times10^{-4}$ & 0.0085 & $6.418\times 10^{-7}$ \\
GRB011121 & $7.97\pm2.2$ & 0.36 & $1.104\times 10^{-11}$ \\
GRB031203 & $(8.2\pm3.5)\times10^{-3}$ & 0.105 & $1.147\times 10^{-7}$ \\
GRB050126 & $[0.4-3.5]$ & 1.29 & $2.664\times 10^{-12}$ \\
GRB050223 & $(8.8\pm4.4)\times10^{-3}$ & 0.5915 & $1.575\times 10^{-11}$ \\
GRB050525A & $2.3\pm0.5$ & 0.606 & $1.482\times 10^{-11}$ \\
GRB050801 & $[0.27-0.74]$ & 1.38 & $2.330\times 10^{-12}$ \\
GRB050826 & $[0.023-0.249]$ & 0.297 & $9.790\times 10^{-11}$ \\
GRB051006 & $[0.9-4.3]$ & 1.059 & $4.018\times 10^{-12}$ \\
GRB051109B & -- & 0.08 & $4.243\times 10^{-9}$ \\
GRB051117B & $[0.034-0.044]$ & 0.481 & $2.674\times 10^{-11}$ \\
GRB060218 & $(5.4\pm0.54)\times10^{-3}$ & 0.0331 & $5.790\times 10^{-8}$ \\
GRB060505 & $(3.9\pm0.9)\times10^{-3}$ & 0.089 & $3.102\times 10^{-9}$ \\
GRB060614 & $0.22\pm0.09$ & 0.125 & $1.150\times 10^{-9}$ \\
GRB060912A & $[0.80-1.42]$ & 0.937 & $5.259\times 10^{-12}$ \\
GRB061021 & -- & 0.3463 & $6.424\times 10^{-11}$ \\
GRB061110A & $[0.35-0.97]$ & 0.758 & $8.583\times 10^{-12}$ \\
GRB070419A & $[0.20-0.87]$ & 0.97 & $4.868\times 10^{-12}$ \\
GRB071112C & -- & 0.823 & $7.073\times 10^{-12}$ \\
GRB081007 & $0.18\pm0.02$ & 0.5295 & $2.086\times 10^{-11}$ \\
GRB090417B & $[0.17-0.35]$ & 0.345 & $6.490\times 10^{-11}$ \\
GRB090814A & $[0.21-0.58]$ & 0.696 & $1.054\times 10^{-11}$ \\
GRB100316D & $(6.9\pm1.7)\times10^{-3}$ & 0.059 & $1.042\times 10^{-8}$ \\
GRB100418A & $[0.06-0.15]$ & 0.6235 & $1.380\times 10^{-11}$ \\
GRB101225A & $[0.68-1.2]$ & 0.847 & $6.617\times 10^{-12}$ \\
GRB110106B & $0.73\pm0.07$ & 0.618 & $1.411\times 10^{-11}$ \\
GRB120422A & $[0.016-0.032]$ & 0.283 & $1.119\times 10^{-10}$ \\
GRB120714B & $0.08\pm0.02$ & 0.3984 & $4.400\times 10^{-11}$ \\
GRB120722A & $[0.51-1.22]$ & 0.9586 & $4.998\times 10^{-12}$ \\
GRB120729A & $[0.80-2.0]$ & 0.8 & $7.557\times 10^{-12}$ \\
GRB130511A & -- & 1.3033 & $2.609\times 10^{-12}$ \\
GRB130831A & $1.16\pm0.12$ & 0.4791 & $2.702\times 10^{-11}$ \\
GRB140318A & -- & 1.02 & $4.358\times 10^{-12}$ \\
GRB140710A & -- & 0.558 & $1.825\times 10^{-11}$ \\
GRB150727A & -- & 0.313 & $8.471\times 10^{-11}$ \\
GRB150821A & $15.37\pm3.86$ & 0.755 & $8.664\times 10^{-12}$ \\
GRB151029A & $0.44\pm0.08$ & 1.423 & $2.195\times 10^{-12}$ \\
GRB151031A & -- & 1.167 & $3.270\times 10^{-12}$ \\
GRB160117B & -- & 0.87 & $6.222\times 10^{-12}$ \\
GRB160425A & -- & 0.555 & $1.850\times 10^{-11}$ \\
GRB161129A & $1.3\pm0.2$ & 0.645 & $1.269\times 10^{-11}$ \\ \bottomrule
\end{tabular} 
\caption{Sample of 41 long GRBs taken from \citet{der17}. For each of them, we indicate their basic properties, and the rate we calculated following the method outlined in Section \ref{sec:results}.}
\label{tab:2}
\end{table}

The sample consists of events detected between 1998 and 2016. Over such a large time span, several instruments were used to perform the detection of each burst, including BeppoSAX \citep[GRBM,][]{boe97}, INTEGRAL \citep[IBAS,][]{mer03}, and the Neil Gehrels Swift Observatory \citep[BAT,][]{geh04}. We list in Table \ref{tab:1} the effective time span for each instrument together with the field of view of its gamma-ray detector. To date, this is the most recent complete sample published for LLGRBs. Since we take into consideration a time-averaged rate of LLGRBs within the detection volume in our method, the addition of more recent GRBs would be balanced by an increase of the operation period, making the final rate value nearly constant.

The variation of the field of view is a bias of this sample, which we deconvolve by taking into account the fact that the GRBs are isotropically distributed. The fact that we do not consider more recent LLGRBs, such as those detected by Fermi \citep[GBM,][]{mee09}, also contribute to the reduction of this bias, as we would be introducing variations in the size of the fields of view. Most of the bursts were detected by the Neil Gehrels Swift Observatory (38 of them), with BeppoSAX performing two detections and INTEGRAL only one.

\begin{table}
\begin{tabular}{@{}lccc@{}}
\toprule
Instrument & FOV [sr] & Operation period [yr] & Considered years of operation \\ \midrule
BeppoSAX & $4\pi$ & 7 &  1996$-$2002 \\
Integral & 0.02 & 14 &  2002$-$2016 \\
Swift & 1.4 & 12 & 2004$-$2016\\ \bottomrule
\end{tabular}
\caption{Summary of instruments used to detect GRBs in our sample.}
\label{tab:1}
\end{table}

In this work, we only consider UHECRs composed of baryons. For the purposes of our study, we require a point of comparison between our predicted UHECR flux and the observed UHECR flux. We have used the results of the Pierre Auger Observatory \citep{abr04}, where the canonical UHECR flux is obtained by numerically integrating the cosmic ray energy spectrum over the ranges for UHECRs above $10^{19}$ eV. The most recent estimate for this value is approximately 1 particle km$^{-2}$ yr$^{-1}$ \citep[e.g.][]{pac17, aab20}.

\section{Methods}
\label{sec:results}
We follow the method of \citet{gue07} and \citet{how14} to compute the expected rates of LLGRBs, like GRB980425, in the local universe. We calculate the rates as: 

\begin{equation}
    R_\text{GRB} = \lp\frac{V_\text{max}}{\text{Mpc}^3}\rp^{-1} \lp\frac{\Omega}{4\pi \:\text{sr}}\rp^{-1} \lp\frac{T}{\text{yr}}\rp^{-1}.
\end{equation}

In this equation, $\Omega$ represents the sky coverage, as listed in Table \ref{tab:1}. $T$ is the time elapsed between the mission launch and the most recent GRB in our sample, as listed in Table \ref{tab:1}. We do not correct for any duty cycle effect, as this is a negligible effect. Finally, $V_\text{max}$ represents the maximum detection volume of each burst:

\begin{equation}
    V_\text{max} = \int_0^{z_\text{max}} \frac{dV}{dz}dz; \qquad \frac{dV}{dz} = \frac{4\pi c}{H_0} \frac{d_L^2(z)}{(1+z)^2 h(z)},
\end{equation}

\noindent
where $dV/dz$ refers to the number density of GRBs at a given redshift that remains constant within the Hubble flow. The normalised Hubble flow $h(z)$ is then determined by: 

\begin{equation}
    h(z) \equiv \frac{H(z)}{H_0} = \sqrt{\Omega_M (1+z)^3 + \Omega_\Lambda}.
\end{equation}

 Hence, the expected rate of each GRB can be found using the redshifts listed in Table \ref{tab:2}. Note that Table \ref{tab:2} presents the computed rate for each burst. 

 We also calculate the total expected rate of GRBs $R_\text{tot}$ as the sum of the individual GRB rates $R_\text{GRB}$,
 
 \begin{equation}
     R_\text{tot} = \sum_n R_\text{GRB},
 \end{equation}
 \noindent
 to be $8.340 \times 10^{-7}$ Mpc$^{-3}$ yr$^{-1}$. Therefore, the rate of GRBs within the GZK horizon (which we assume to be $d_\text{GZK} = $ 50 Mpc) \citep[e.g.][]{abb08} that are directed face on is 

\begin{equation}
    R_\text{tot, GZK} =\frac{4}{3} \pi \lp d_\text{GZK}\rp^3 \times R_\text{tot} = 0.437 \: \text{yr}^{-1}.
\end{equation}

By definition, we can only observe face on events. Therefore, we need to provide a beaming correction, $f_b$, to account for the population missed that are not directed face on. Then, under the assumption that GRBs emit $N$ UHECR particles, we can determine the particle luminosity as

\begin{equation}
   L_\text{particle} = N \times R_\text{tot, GZK} \times f_b .
\end{equation}

% \noindent 
We can finally compute the particle flux on Earth from sub-GZK GRBs as:

\begin{equation}
\label{eq:7}
    F_\text{particle} = \frac{L_\text{particle}}{S_\text{GZK}} = \frac{N \times R_\text{tot, GZK} \times f_b}{S_\text{GZK}},
\end{equation}

where $S_\text{GZK}$ is the surface area of the GZK sphere.

\section{The Nature of LLGRBs}
\label{sec:llgrb}

There is currently no clear consensus on the nature of LLGRBs, and several explanations have been invoked in order to explain their origin and properties. The detection of rare Type Ic supernovae with these bursts suggests a common origin with that of typical long GRBs \citep[for instance, compare GRB980425 and GRB130427A:][]{gal98, xu13}, hence the difference may lie with either the properties of the jet or the nature of the central engine. One popular hypothesis is that LLGRBs are powered by a shock breakout, either from the progenitor star itself \citep[e.g.][]{cam06, wan07b, nak12} or from an extended low-mass stellar envelope \citep{nak15}. However, some LLGRBs do not support this argument \citep[e.g. GRB120422A,][]{zha12}. Instead, others have proposed that perhaps the cause lies with the central engine allowing for only low bulk Lorentz factors of the flow that results in the observed low emissivity of the jet \citep{lia10, irw16}, or indeed it may simply be due to a viewing angle effect \citep{zha04}. However, in the study by \citet{dai07}, it was shown that the viewing angle effect cannot explain the dim afterglows of these bursts, and the probability of detecting off-axis GRB980425-like GRBs should be rather low. Thus, the detection of other LLGRBs as well as their less-energetic afterglows supports the notion that these bursts are powered by either a mildly relativistic/energetic outflow, or by shock breakout. 

Regardless of how LLGRBs are theoretically modelled, comparisons with the local Type Ib/c supernova rate indicate that they are not strongly beamed \citep[as discussed in][]{lia07}. This is further compounded with the lack of detection of a jet break in the afterglow of these bursts \citep[e.g. GRB060218 and GRB201015A:][]{sod06, pat23}; suggesting a quasi-spherical nature of the jet. With such large jet opening angles, the beaming correction for these bursts should be less than that of typical long GRBs.

On the other hand, with no certainty on the true beaming of these event, it is more conservative to use a confidence interval for $f_b$ rather than a given value.

This leads us to the question: \textit{what is a good value for $f_b$ in our calculations?} As previously discussed, some of these bursts appear to display properties suggesting quasi-spherical emission, thereby a correct lower boundary is to simply assume $f_b = 1$, indicating no beaming. In such a case, we can compute a theoretical particle flux from Equation \ref{eq:7} as $F_\text{particle} = N \times 8.8 \times 10^{-21}$ particles km$^{-2}$ yr$^{-1}$ mol$^{-1}$. On the other hand, we 
know for sure that we do observe some events, and thus that the beaming angle has to be larger than the one sustained by the Earth surface, hence $f_b \leq S_\text{GZK} / S_\text{Earth}$. Naturally, such a beaming factor is nonphysical, as it vastly over-predicts the number of LLGRBs with respect to the local supernova rate; nonetheless, it is useful as an extreme upper bound as we show here. Then, the particle flux becomes $F_\text{particle} = N \times 5.2 \times 10^{14}$ particles km$^{-2}$ yr$^{-1}$ mol$^{-1}$. We can thus formulate an inequality to express where the true flux lies:\\
\\
$N \times 8.8 \times 10^{-21} \leq F_\text{particle} \leq N \times 5.2 \times 10^{14}$ particles km$^{-2}$ yr$^{-1}$ mol$^{-1}$.
\\
\\
Finally, it is possible to constrain the value for $N$ within the particle flux inequality by using the observed Auger UHECR particle flux. If we express this constraint as the amount of mass accelerated within the jet then, under the assumption that 1 M$_\odot$ contains $\sim 10^{33}$ mol of baryons, in order to reach the observed 1 particle km$^{-2}$ yr$^{-1}$, we require

\begin{equation}
\label{eq_final}
    {10^{-48}} \leq M \leq 10^{-13} M_\odot.
\end{equation}

\section{Discussion}
\label{sec:discussion}
\subsection{Rates}

One may be tempted to compare our rate of LLGRBs with others published in the literature. However, we emphasise that we are using a dedicated sample that has been selected from their afterglow properties, rather than from the prompt properties as in, e.g. \citet{vir09, sun15}. Indeed, it has been discussed numerous times that gamma-ray prompt properties are plagued by selection effects (such as the $T_{90}$ value which is different for each detector; or the end time of the prompt phase), and that X-ray observations are better suited for measuring physical events \citep[e.g. the end of the prompt phase,][]{str13}. Another example is the classification of GRB211211A, which is a "long short GRB": its prompt properties classify it as a long collapsar event, while its global properties indicate that it is a short event linked to a binary merger of compact objects \citep{ras22}. For these reasons, by construction, our rates may differ from other published ones, as they are supposed to be less biased by instrumental effects.

Nonetheless, our observed rate of $8.340\times 10^{-7}$ Mpc$^{-3}$ yr$^{-1}$ is still highly compatible with the rate of Type Ib/c supernovae, assuming a realistic beaming factor. Estimates place the local ($d \lesssim 100$ Mpc) rate of these supernovae at $\sim 9_{-5}^{+3} \times 10^{-6}$ Mpc$^{-3}$ yr$^{-1}$ \citep{cap99, dah04, gue07}, placing our LLGRB rate as $\sim 0.9 \%$.

\subsection{GZK Horizon}

The first obvious suppression mechanism of our calculated UHECR output is the GZK horizon. The predominant energy loss mechanism for UHECRs travelling through the extra-galactic environment is interactions with interstellar radiation fields, in particular the CMB. It is easy to reject potential sources of UHECRs simply due to their existence out of the GZK horizon, which can vary from 50 Mpc to hundreds of Mpc. Out of the GRB population used in this study, only one burst is located within the horizon: GRB980425. It is not to be overlooked that the GRB rates and hence the UHECR flux calculated are primarily driven by that one event, as can be seen by its rate in Table \ref{tab:2}. In considering that burst alone using the rate data in Table \ref{tab:2}, we derive a rate of 0.336 GRB980425-like bursts per year, yielding an UHECR flux of $N \times 6.8 \times 10^{-21} \leq F_\text{particle} \leq N \times 4 \times 10^{14}$ particles km$^{-2}$ yr$^{-1}$ mol$^{-1}$. Assuming all the other events would be suppressed due to the horizon, this corresponds to a suppression of only $\sim$33\%.

The possibility of GRB980425 being a source of UHECRs has also been explored by \citet{mir22b,mir22}. In these studies, the UHECRs produced by the burst were deflected by the IGMF, and subsequently cascaded into secondary gamma rays. Their main argument is that depending on the strength of the IGMF, one would expect UHECRs arriving from this source within the next 100 years. Thus, if LLGRBs are indeed the sources of UHECRs as our calculations have shown, one would expect that historical 980425-like GRBs may explain current observations of the UHECR flux.

\subsection{Magnetic Fields}

A second suppression mechanism is the role of Galactic and Intergalactic Magnetic Fields (GMF and IGMF respectively) on the trajectory and arrival time of UHECR particles. While our understanding of UHECR propagation within the GMF has increased over the decades \citep[see e.g.][]{tan98, tin05, far12}, the IGMF is less understood, and can vary in both strength and coherence length by several orders of magnitude. At first glance, the GMF dipole can indeed introduce a potential break into the isotropy hypothesis of the common $\Lambda$-CDM model. However, magnetic fields cannot introduce anisotropies \citep{eic20}. As gamma-ray bursts are isotropically distributed, we do not expect the GMF to affect the results. The situation is less clear for the IGMF. Numerical simulations have been performed to understand the propagation and deflection of UHECRs in the IGMF, indicating that the charge and rigidity of the cosmic rays play an important role in their behaviour \citep[e.g.][]{erd16, hac18, mag19}. Given the uncertainties in the IGMF, one could naively suppose that the magnetic deflections are responsible for a large dilution in the UHECR fluxes produced by the GRBs. It is that dilution which would cause the very low detection rate observed on Earth, despite a larger production rate at the acceleration sites.

On the other hand, the IGMF is unorganised enough so that we can consider that the same amount of particles deflected from us will also deflected toward us from other directions, recalling that the distribution of GRBs is isotropic \citep{mee92}. The IGMF thus should not interfere in the suppression of the flux, but instead with the arrival time. Even so, as noted in the case for GRB980425, the arrival times should be delayed by hundred of years for sub-GZK events. As for the GMF, there is no observational signature of the dipole in cosmic ray arrival directions \citep{aab15}, thus again the same previous argument should apply.

\section{Conclusion}
\label{sec:conclusion}

In this work, we explored the theoretical contribution of the UHECR flux by GRBs. We concentrated on the contribution of the low-luminosity bursts, which form the bulk of the nearby population of GRBs. We found that in order to reconcile our theoretical flux with the observed flux, these bursts should accelerate at most $10^{-13}$ M$_\odot$ of ultra-high energy cosmic rays. 

Such a hypothesis can be tested by studying the composition of the UHECRs and comparing it with the composition of the surrounding medium of GRBs obtained through spectroscopy. In optical, current experiments like X-Shooter already gives access to the composition of the host galaxy a few days after the event. However, one may expect a medium rich in metals expelled by the stellar progenitor before the burst. In that case, the Athena+ instruments \citep{barr23} could perform a study of such environments in X-rays.

As for the particle experiments, an excess of heavy nuclei (compared to simple protons) for the highest energies would then confirm the electromagnetic observations. It is therefore very important that particle observatories and large spectroscopic missions are able to work together during the next decade on those topics.

Lastly, improving neutrino detectors is also a top priority to solve this issue. Neutrinos, being by-products of hadronic interactions, and the lack of their detection in association with gamma-ray bursts (in particular the prompt emission) is indicative of a lack of a significant proportion of protons into the jet. Proving the detection of an absence (and not an absence of detection) is by definition a tricky task, but future more powerful neutrino detectors could for sure be able to do so.

\section*{Acknowledgements}

The authors thank the anonymous referee for their useful comments and discussion on this paper. This research was supported by the Australian Research Council Centre of Excellence for Gravitational Wave Discovery (OzGrav), through project number CE170100004. E.M. acknowledges support support from the Zadko Postgraduate Fellowship and International Space Centre (ISC). We gratefully acknowledge support through NASA-EPSCoR grant NNX13AD28A. N.B.O. also acknowledges financial support from NASA-MIRO grant NNX15AP95A, NASA-RID grant NNX16AL44A, and NSF-EiR grant 1901296. F.P. was supported by funding from the Forrest Research Foundation. 

\section*{Data availability}
The data underlying this article will be shared on reasonable request to the corresponding author.

\bibliographystyle{mnras}
\bibliography{bibliography_emoore}
% Don't change these lineables
\bsp	% typesetting comment
\label{lastpage}
\end{document}